\documentclass[12pt,preprint]{aastex}

\newcommand{\Rmnum}[1]{\expandafter\@slowromancap\romannumeral #1@}

\shorttitle{Photospheric Magnetic and Current Helicities and Subsurface Kinetic Helicities}
\shortauthors{Seligman, Petrie and Komm}

\begin{document}


\title{A Combined Study of Photospheric Magnetic and Current Helicities and Subsurface Kinetic Helicities of Solar Active Regions during 2006-2013}


\author{D. Seligman$^1$, G.J.D. Petrie$^2$ \& R. Komm$^2$}
\affil{$^1$University of Pennsylvania, USA; National Solar Observatory REU Program\\
$^2$National Solar Observatory, Tucson, AZ 85719, USA}


%\altaffiltext{1}{} 
%\altaffiltext{2}{}

%%%%%%%%%%%%%%%%%%%%%%%%%%%%%%%%%%%%%%%%%%%%%%%%%%%%%%%%%%%%%%%%%%%%%%%%%%%%%%%%%
\begin{abstract}

We compare the average photospheric current helicity $H_c$, photospheric twist parameter $\alpha$ (a well-known proxy for the full relative magnetic helicity), and subsurface kinetic helicity $H_k$ for 194 active regions observed between 2006-2013. We use 2440 Hinode photospheric vector magnetograms, and the corresponding subsurface fluid velocity data derived from GONG (2006-2012) and HMI (2010-2013) Dopplergrams. We find a significant hemispheric bias in all three parameters. The subsurface kinetic helicity is preferentially positive in the southern hemisphere and negative in the northern hemisphere. The photospheric current helicity and the $\alpha$ parameter have the same bias for strong fields ($|B|>1000$~G) and no significant bias for weak fields (100~G~$<|B|<500$~G). We find no significant region-by-region correlation between the subsurface kinetic helicity and either the strong-field current helicity or $\alpha$. Subsurface fluid motions of a given handedness correspond to photospheric helicities of both signs in approximately equal numbers. However, common variations appear in annual averages of these quantities over all regions. Furthermore, in a subset of 77 regions we find significant correlations between the temporal profiles of the subsurface and photospheric helicities. In these cases, the sign of the linear correlation coefficient matches the sign relationship between the helicities, indicating that the photospheric magnetic field twist is sensitive to the twisting motions below the surface.

\end{abstract}
\keywords{solar magnetic fields, solar photosphere, solar corona}

%%%%%%%%%%%%%%%%%%%%%%%%%%%%%%%%%%%%%%%%%%%%%%%%%%%%%%%%%%%%%%%%%%%%%%%%%%%%%

\section{Introduction}
\label{sect:intro}

A solar active region is an area in the photosphere characterized by an intense magnetic field. The most spectacular and dangerous solar phenomena, flares and coronal mass ejections, derive from twisted active region fields in the magnetically dominated atmosphere. These fields are created by dynamo processes in the fluid-dominated interior (e.g., Charbonneau~2010). The differential rotation inside the Sun is believed to create a shear flow at active latitudes which stretches the field out and winds it around the rotation axis, creating very strong toroidal fields at active latitudes. This dynamo process is known as the $\Omega$-effect (Parker~1955). These strong toroidal fields then rise buoyantly and become subject to a Coriolis force of opposite handedness in each hemisphere (Fan~2009). The fields may therefore twist before their arrival at the surface, but there are differing theoretical predictions of the sense of this twist.

If a vector field describes a clockwise/anti-clockwise screwing motion along the direction of the flux associated with this field, then the field is defined to have right-/left-handed sense or twist. In the case of upflows in the solar interior subject to the Coriolis force, a right-/left-handed upflow would be observed from above to describe an anti-clockwise/clockwise motion. Following the discussion in Longcope et al.~(1998), fluid motions with a right-handed sense will deform a flux tube in a left-handed sense (Moffatt~1978). The Coriolis force acting on an upward flow in the southern hemisphere of the Sun would produce motions of this right-handedness. We will mention two effects such fluid motions may have on a field: the $\alpha$-effect (Parker~1955) and the $\Sigma$-effect (Longcope et al.~1998). In dynamo theory the term $\alpha$-effect is often used to refer to the general conversion of a toroidal to a poloidal field, but here, we refer to the rising and then tilting of the flux tube toward the equator due to the Coriolis force (Parker~1955), which induces a twist of the magnetic field lines. The $\alpha$-effect and the $\Sigma$-effect differ in the following way. The $\alpha$-effect relates the deformation of any field to its associated fluid velocity via the standard induction equation of magnetohydrodynamics. The $\Sigma$-effect applies only to isolated flux tubes, relating the deformation of the field inside the tube to the fluid velocity outside the tube. Right-handed motions would produce a left-handed twist in the field frozen into the fluid. However, right-handed motions acting on an isolated flux tube would produce a left-handed ``writhe'' of the tube, a left-handed deformation of the tube axis. This would lead, by conservation of magnetic helicity, to right-handed twist (Longcope et al.~1998). Thus the $\Sigma$-effect should produce $J_z/B_z > 0$ in the southern hemisphere and $J_z/B_z < 0$ in the northern hemisphere, consistent with the observed hemispheric helicity biases discussed in the following.

The $\alpha$-effect plays a central role in solar dynamo models (Charbonneau~2010), where it not only twists but also amplifies the field, e.g., a shearing motion associated with interior differential rotation may amplify the field into strong flux ropes which then buoyantly rise, and twist in response to the Coriolis force. The $\Sigma$-effect does not change the flux of the flux tube which is constant by definition. The $\Sigma$-effect therefore cannot explain the generation of the flux, but it may restructure the buoyantly rising flux that is generated in the interior, possibly by the $\alpha$-effect. On the other hand, the $\alpha$-effect cannot explain the observed handedness of the emerged photospheric field.

The aim of this project is to investigate the nature of the dynamo process by comparing observed subsurface kinetic helicities and photospheric magnetic and current helicities of active-region fields. The relative magnetic helicity is likely to be approximately conserved in the solar atmosphere  (Berger~1984) on timescales of an active region's observed passage across the solar disk (1-2 weeks), whereas the helicity proxies that we will describe are not conserved quantities. Since relative helicities are difficult to extract from observations, we confine ourselves to studying proxies that are much simpler to derive observationally, and that characterize the handedness of the field as it crosses the photospheric layer from where the field measurements originate.

For many years it has been known that the solar magnetic field has a hemispheric bias in its magnetic helicity or twist. A hemispheric preference for positive/negative twist in the southern/northern hemisphere was seen in chromospheric H$_{\alpha}$ images by Hale~(1927) and Richardson~(1941). The bias was initially believed to be caused by the Coriolis force, analogous to terrestrial hurricanes, but the twist of the low-$\beta$ fields of the chromospheric and higher atmospheric layers is now known to be associated with near-force-free electric currents. However, the ultimate origin of these currents may yet be the Coriolis force acting on interior plasma flows.

The twist of the observed photospheric magnetic vector field can be characterized by computing the ``force-free'' parameter $\alpha$ (not to be confused with the $\alpha$-effect described above),

\begin{equation}
\alpha = \frac{(\nabla\times{\bf B})_z}{B_z} = \frac{4\pi J_z}{B_z}
\label{eq:alpha}
\end{equation}

\noindent where $J_z$ is the vertical component of the electric current density. Although the photospheric field is not generally force-free (e.g., Metcalf et al.~1995), the $\alpha$ parameter nevertheless offers a useful estimate of the twist of the field that penetrates the thin photospheric layer of the solar atmosphere. Photospheric vector magnetogram observations have shown a preference of $\alpha$ to take positive/negative values, corresponding to right-/left-handed twist, in the southern/northern hemisphere (Pevtsov et al.~1994, 1995).

A related quantity is the current helicity (Seehafer~1990). The vertical component of the current helicity density,

\begin{equation}
H_c = (\nabla\times{\bf B})_z B_z = 4\pi J_zB_z,
\label{eq:Hc}
\end{equation}

\noindent shows a similar preference for positive/negative values in the southern/northern hemisphere (Seehafer~1990, Gosain et al.~2013). Only the vertical component of the current helicity density can be derived from vector magnetograms measuring the field in a single atmospheric layer. Fortunately, this component does offer a useful proxy for the helicity of the currents crossing this layer and is therefore useful to our study. The two parameters are related to each other by the formula $H_c = \alpha B_z^2$ (Hagyard \& Pevtsov~1999). The $\alpha$ and $H_c$ parameters have been used by many authors (e.g., Seehafer~1990, Pevtsov et al.~1994, 1995, Gosain et al.~2013) as a proxy for the magnetic helicity, which is difficult to calculate in practice. In this paper we will use these parameters in a similar way.

Much observational analysis of the hemispheric helicity bias has been reported. In the first attempt to quantify the current helicity in a general way, Seehafer~(1990) concluded from best-fitting force-free models of atmospheric structures that the current helicity was predominantly negative/positive in the northern/southern hemisphere. Pevtsov et al.~(1994) extended this work by analyzing 46 vector magnetograms from the Stokes Polarimeter of Mees Solar Observatory and Pevtsov et al.~(1995) used a combination of vector magnetograms and published results to study the average best-fitting $\alpha$ value of 69 active regions. Consistent with the results of Seehafer~(1990), Pevtsov et al.~(1994, 1995) found that a significant majority of the active regions in the northern/southern hemisphere had negative/positive helicity. This hemispheric helicity rule was also supported by the results of Abramenko et al.~(1997), Bao \& Zhang~(1998), Pevtsov et al.~(2001, 2008), Hagino \& Sakurai~(2004) and Zhang et al.~(2010) among others.

However, Zhang et al.'s~(2010) observed butterfly (latitude-time) plots of $H_c$ and $\alpha$ suggested more complex, cycle-dependent behavior: these plots featured inversions of sign at the beginnings and ends of the butterfly wings, corresponding to the beginnings and ends of activity cycles. Some other studies have also revealed signs of cycle-dependent helicity patterns, as well as differences in behavior between weak and strong fields. Using 17200 vector magnetograms observed by Huairou Solar Observing Station of the Chinese National Astronomical Observatory from 1997-2004, Zhang~(2006) found the usual hemispheric biases (negative/positive in the northern/southern hemisphere) in $\alpha$ and $H_c$ for the weak fields (100~G~$<|B_z|<500$~G) but an opposite bias in the strong fields ($|B_z|>1000$~G), a result confirmed by Hao \& Zhang~(2011) using Hinode vector magnetograms. Hao \& Zhang also found that the 34 active regions from early cycle 24 in their sample followed the hemispheric helicity rule, whereas the 30 active regions from late cycle 23 did not. Using SOLIS/VSM vector synoptic maps covering the period March 2011 - December 2012 (early cycle 24), Gosain et al.~(2013) found that the strong fields exhibited the usual hemispheric bias and the weak fields exhibited a weak bias of opposite sign. Komm et al.~(2014a) found that, in general, active regions follow the Seehafer-Pevtsov bias, but two regions observed to have a ``whirly'' magnetic structure (i.e., the chromospheric fibrils spiral out from the umbra in generally the same direction) indicating magnetic helicity, exhibited the opposite bias. It is clear that, while a persistent helicity bias exists through much of the cycle, significant populations of active region fields do not follow this bias.

Comparisons of the magnetic twist and current helicity in the photosphere to the kinetic helicity of the corresponding plasma flows in the interior might enable us to characterize the transport of helicity from the interior to the atmosphere. Towards this goal, we compare the photospheric twist parameter $\alpha$ and the photospheric vertical current helicity component $H_c$ to the subsurface vertical kinetic helicity component. We expect the kinetic helicity of the subsurface flows to be related to the twist parameters associated with the photospheric fields because the magnetic field and fluid flow vectors are coupled via nearly ideal magnetohydrodynamical processes of highly conducting plasmas whose fields and plasmas are frozen together (e.g., Priest~1982): in high-$\beta$ regimes (plasma pressure \textgreater\ magnetic pressure) characteristic of the solar interior the fields are advected and twisted by the flows whereas in low-$\beta$ regimes (plasma pressure \textless\ magnetic pressure) characteristic of the solar atmosphere the fluid flows are constrained by the structure of the field. The main source of solar photospheric magnetic and current helicity is therefore widely believed to be the kinetic helicity of subsurface flows.

Past authors have also found hemispheric biases in the helicity of the corresponding subsurface flows (Zhao~2004). The product of the divergence and the curl of the horizontal flows, a proxy for the kinetic helicity density, has been used to represent the kinetic helicity of the subsurface flows. The vertical contribution to the kinetic helicity is,

\begin{equation}
H_k = w_z v_z,
\label{eq:Hk}
\end{equation}

\noindent where $v_z$ is the vertical component of the velocity vector ${\bf v}$, and $\omega_z$ is the vertical component of the vorticity vector ${\bf \omega}={\bf\nabla}\times{\bf v}$. The vertical vorticity $\omega_z$ is the vorticity associated with the horizontal velocity components. Equation~(\ref{eq:Hk}) is directly analogous to the vertical component of the vertical current helicity defined in Equation~(\ref{eq:Hc}). $H_k$ has been found to take preferentially positive/negative values in the southern/northern hemisphere (Komm et al.~2007). This hemispheric bias is most likely due to the Coriolis force (Brun et al.~2004, Egorov et al.~2004), but the bias is not strict. Komm et al.~(2014b) found that the average kinetic helicity density at all depths from 2-7~Mm follow this same hemispheric rule with positive/negative values in the southern/northern hemisphere. Though only 55\% of locations were found to follow this rule, these locations had larger helicity values than the locations that did not follow the rule.

Since the three helicity parameters $H_k$, $H_c$ and $\alpha$, show similar hemispheric biases, they may be physically related. For example, the $\Sigma$-effect is consistent with subsurface kinetic helicities and photospheric magnetic and current helicities having the same sign. However, according to past work, the relationship between the subsurface flows and the photospheric magnetic field appears not to be simple. Gao et al.~(2009) found typical hemispheric biases in $\alpha$ and $H_c$ of 38 active regions but no such pattern in the corresponding kinetic helicities in two depth ranges, 0-3~Mm and 9-12~Mm, and Maurya et al.~(2011) found that, while the vertical kinetic and current helicities showed similar hemispheric biases, no significant correlation between these parameters was apparent. Gao et al.~(2012) studied the temporal variation of $H_k$ and $H_c$ (weighted by flow speed and field strength, respectively) and found that the two parameters' temporal variations were reasonably well correlated even though the parameters generally did not have the same sign as each other. They used this latter result to explain why, when one takes snapshots (or, as in the present paper, temporal averages) of the helicity parameters for each region, then these results may not be correlated on a region-by-region basis even when the helicity parameters show common temporal variations.

We will use a combination of Hinode vector magnetograms and helioseismic data from the NSO's Global Oscillation Network Group (GONG) network and from Helioseismic and Magnetic Imager (HMI) on NASA's Solar Dynamics Observatory (SDO) satellite. Maurya et al.~(2011) used a similar combination of GONG Doppler images and Hinode vector magnetograms, covering 91 cycle 23 active regions, and they
%used different Stokes inversion and horizontal-field $180^{\circ}$ disambiguation techniques from the ones applied to the Hinode data used in this paper, and
represented each active region with a single vector magnetogram. Gao et al.~(2009) used MDI Dopplergrams and Huairou vector magnetograms covering 38 cycle 23 active regions. We will include in our analysis Hinode/SOT vector magnetograms where available, and matching GONG and HMI helioseismic data, covering 194 active regions between 2006 and 2013. We will compare subsurface kinetic helicities vs. photospheric current helicities and magnetic twists ($H_c$ and $\alpha$) to characterize the global relationship between these quantities.

We have collected vector magnetogram data and helioseismic kinetic helicity data covering the decline of cycle 23 and the rise of cycle 24 (2006-2013). We will plot in time the global patterns of the helicity parameters and determine whether their overall behavior is cycle-dependent as Zhang et al. (2010) and Gosain et al.~(2013) found. We will also explore in detail whether common patterns appear in the photospheric and subsurface helicities, not only in their overall distributions but on a region-by-region basis. The basic goal of our study is to determine whether significant correlations between the photospheric and subsurface helicity patterns exist that point to a physical link between the subsurface flows and the photospheric fields. Such a physical link has long been anticipated by theory (Parker~1955, Longcope et al.~1998, Fan and Gong~2000) but convincing evidence has not been found in the observations. Since the $\alpha$-effect is consistent with opposite-sign relationships between the subsurface and photospheric helicities, whereas the $\Sigma$-effect is consistent with same-sign relationships between the subsurface and photospheric helicities, the study may shed light on which processes are responsible for the twist of the solar magnetic fields that we observe. In the following, we will refer to subsurface and photospheric helicities, where the subsurface helicity is the vertical kinetic helicity $H_k$ defined by Equation~(\ref{eq:Hk}) and the photospheric helicities are the vertical electric current helicity $H_c$ and the twist parameter $\alpha$ defined by Equations~(\ref{eq:alpha},\ref{eq:Hc}).

The paper is organized as follows. We describe the data in Section~\ref{sect:data}, then we present the results for averaged helicity values of all active regions in Section~\ref{sect:arch} and the results for temporal variations of the helicity values for selected active regions in Section~\ref{sect:temporal}. We close with a discussion in Section~\ref{sect:conclusion}.

\section{Data}
\label{sect:data}

We use vector magnetograms observed by the Spectro-polarimeter (SP) on the Hinode satellite (Kosugi et al.~2007). The Hinode satellite was launched in late 2006, catching the end of cycle 23, then covering the cycle 23 minimum and the ascent of cycle 24. The Hinode data are unique in providing continuous high-quality vector magnetogram coverage of the past 7+ years, including a solar cycle transition, with high sensitivity and good spatial resolution. We therefore have an opportunity to explore Zhang et al.'s~(2010) conclusion that hemispheric helicity biases might tend to reverse at the beginnings and ends of activity cycles. Past studies (e.g., Zhang~2006, Hao \& Zhang~2011, Gosain et al.~2013) have identified different patterns of behavior in the helicities of strong and weak magnetic fields. We will therefore compare $H_c$ and $\alpha$ values derived from photospheric magnetograms for a large sample of active regions with helioseismic results for the subsurface kinetic helicities of these regions, treating strong and weak fields separately.

The Hinode/Spectro-Polarimeter obtains line profiles of two magnetically sensitive Fe lines at 630.15 and 630.25~nm and nearby continuum, using a $0.16^{\prime\prime} \times 164^{\prime\prime}$ slit. Of the four mapping modes of operation (normal map, fast map, dynamics, and deep magnetogram, Tsuneta et al.~2008) we only use the normal maps (about 0.16$^{\prime\prime}$~pixel$^{-1}$) and fast maps (about 0.32$^{\prime\prime}$~pixel$^{-1}$). For this study, we have used Level 2 Hinode magnetograms processed with new, faster ambiguity resolution method developed by Rudenko \& Anfinogentov (2013). In this method, the direction of the transverse field is determined using a method based on the principle of minimizing the deviation of the field from the reference potential field. Gosain et al.~(2013) validated the method by comparing with magnetograms disambiguated using the non-potential field calculation (NPFC) technique of Georgoulis~(2005). In this project we analyze 2440 Hinode vector magnetograms encompassing 194 active regions from 2006 to 2013, in conjunction with GONG (2006-2012) and HMI (2010-2013) helioseismic time series for the regions.

We use high-resolution full-disk Doppler data from the GONG network to derive the subsurface flows associated with the regions. We determine the horizontal components of the subsurface flows with a ring-diagram analysis using the dense-pack technique (Haber et al.~2002) adapted to GONG data (Corbard et al.~2003). Each full-disk Doppler image is divided into $15^{\circ}$ overlapping patches whose centers are spaced $7.5^{\circ}$ apart ranging over $\pm 52.5^{\circ}$ in latitude and central meridian distance (CMD). From these we derive daily flow maps of horizontal velocities. We also estimate the vertical velocity component from the convergence of the horizontal flows using mass conservation (Komm et al.~2004, Komm~2007). From the subsurface flows we calculate the kinetic helicity density ${\bf w}\cdot {\bf v}$ from the fluid velocity ${\bf v}$ and vorticity ${\bf w} = {\nabla}\times{\bf v}$. For comparison with $\alpha$ and $H_c$ we focus on the vertical contribution to the kinetic helicity $H_k$, defined by Equation~(\ref{eq:Hk}). We integrate $H_k$ over a suitable range in depth from 2 to 7 Mm as in
Komm et al~(2014b).

In addition, we determine horizontal subsurface flows using HMI 
Dopplergrams processed by the HMI ring-diagram pipeline 
(Bogart et al.~2011a, 2011b) 
and calculate the vertical contribution to the kinetic helicity $H_k$.  
Full documentation on the HMI pipeline analysis modules and associated data 
products can be found on the web pages of the HMI Ring Diagrams Team 
({\url{http://hmi.stanford.edu/teams/rings/}}).  HMI pipeline results are 
available through the Joint Science Operations Center or JSOC ({\url{http://jsoc.stanford.edu}}). 
The centers of the $15^{\circ}$ patches are spaced by $7.5^{\circ}$ in 
latitude and CMD with a range of up to $75^{\circ}$ in Latitude and CMD.  
At latitudes higher than $30^{\circ}$, we had to interpolate the inferred 
flows linearly in CMD on a $7.5^{\circ}$ grid since the patches are spaced sparser at these latitudes.

\section{Averaged Helicities for All Active Regions}
\label{sect:arch}

\begin{figure} 
\begin{center}
%\resizebox{0.99\textwidth}{!}{\includegraphics*{scatter_kin_hc_alpha_strong.ps}}
\resizebox{0.99\textwidth}{!}{\includegraphics*{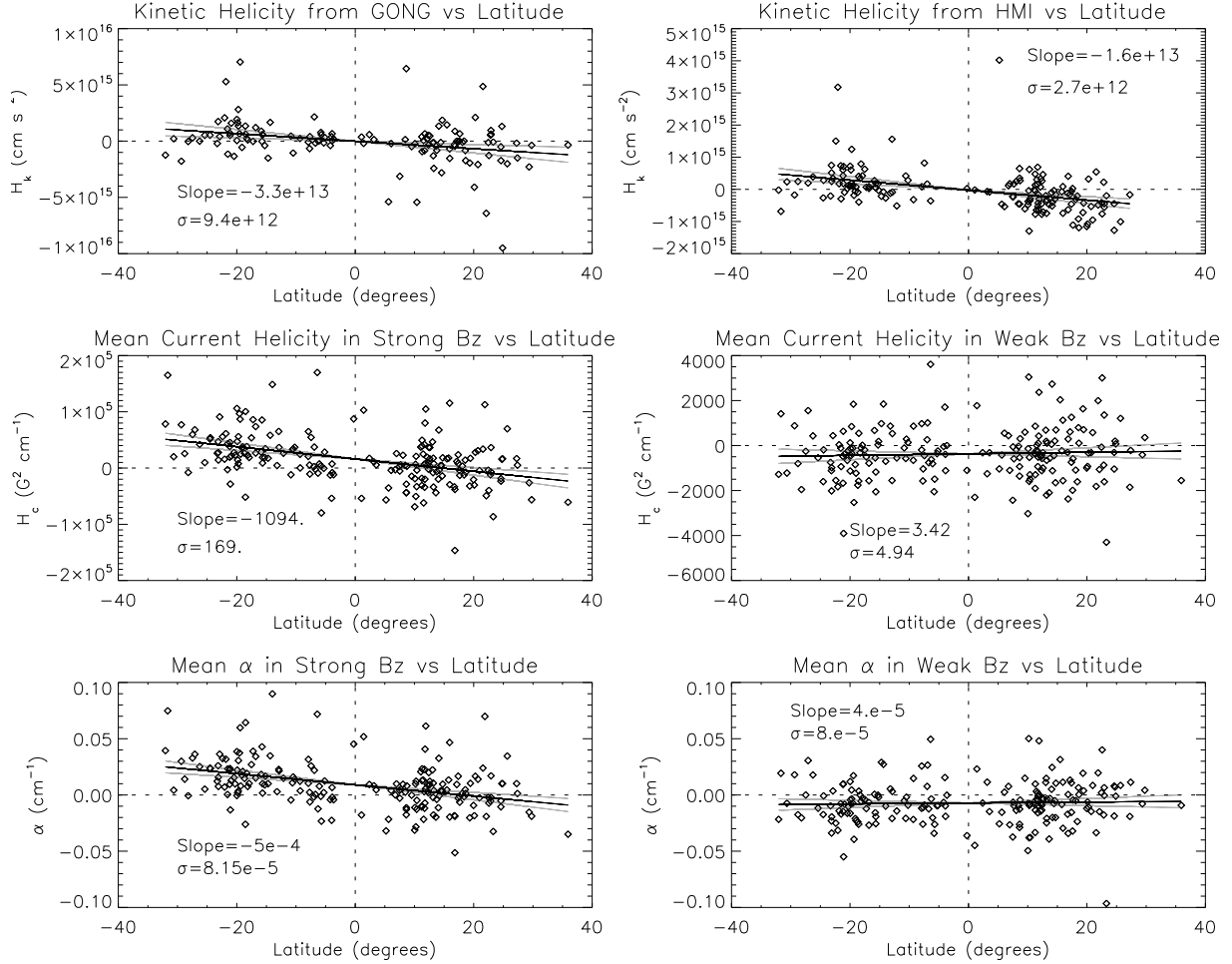}}
\end{center}
\caption{Scatter plots of the subsurface kinetic helicity $H_k$ (top plots), the photospheric current helicity $H_c$ (middle plots), and the photospheric magnetic twist parameter $\alpha$ (bottom plots), all plotted against heliographic latitude. Weak- and strong-field $H_c$ and $\alpha$ values are plotted separately, as are GONG (2006-2012) and HMI (2010-2013) $H_k$ data. Data points of positive/negative latitude are in the northern/southern hemisphere. In each figure, the middle line represents the best fit using a linear regression model and the two grey lines represent  $2\sigma$ error bars. In each plot the values for the slope and $\sigma$ of the best-fit line from the linear regression analysis are quoted. The scaling of the vertical axes is not identical in all plots.}
\label{fig:hel_lat}
\end{figure}

\begin{figure} 
\begin{center}
%\resizebox{0.99\textwidth}{!}{\includegraphics*{masterplot.ps}}
\resizebox{0.99\textwidth}{!}{\includegraphics*{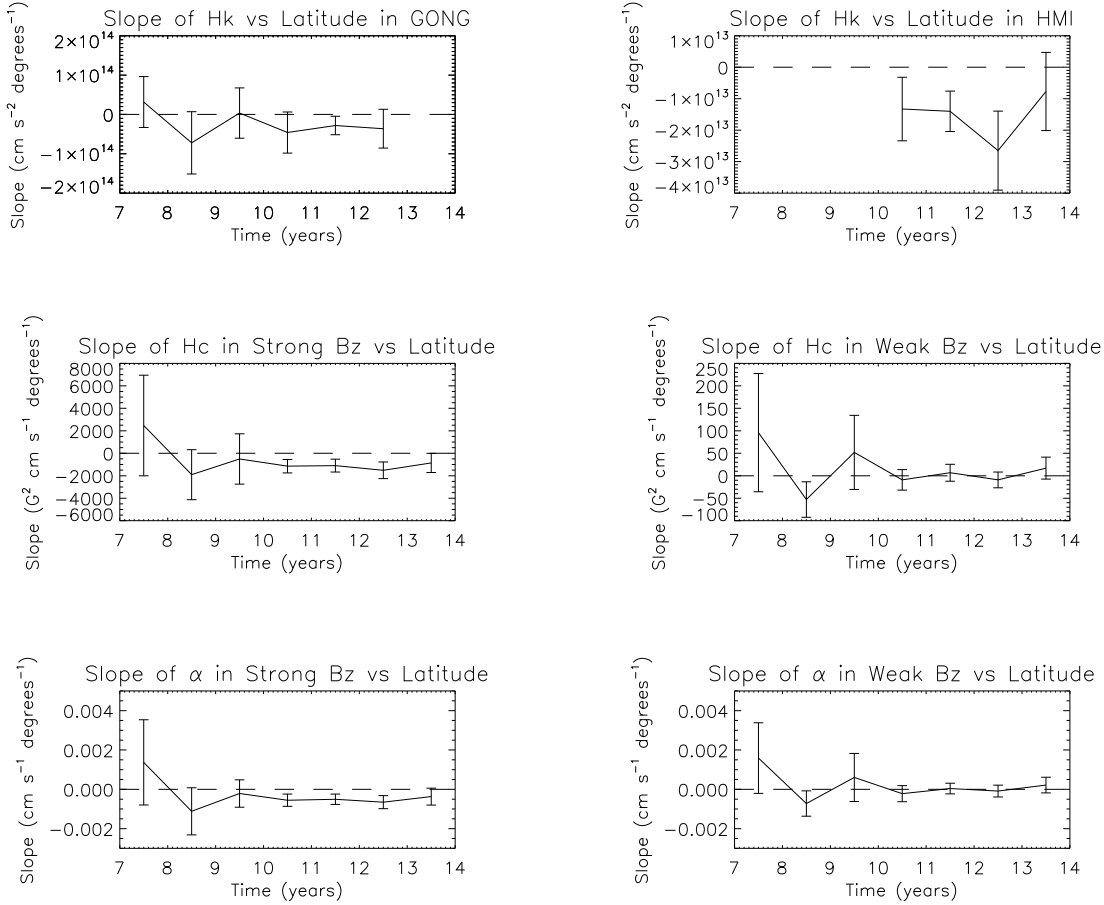}}
\end{center}
\caption{Shown are annual linear regression slopes, i.e., the slopes of the best-fit lines in Figure~\ref{fig:hel_lat}, for the subsurface kinetic helicity $H_k$ (top plots), the photospheric current helicity $H_c$ (middle plots), and the photospheric magnetic twist parameter $\alpha$ (bottom plots), all against heliographic latitude. Slopes for weak- and strong-field $H_c$ and $\alpha$ values are plotted separately, as are slopes for GONG (2007-2012) and HMI (2010-2013) $H_k$ data. Each error bar represents the $2\sigma$ error from the regression analysis. The years labeled on the x-axes denote the years 2007-2014.}
\label{fig:masterplot}
\end{figure}

\begin{figure} 
\begin{center}
%\resizebox{0.99\textwidth}{!}{\includegraphics*{hc_alpha_regress.ps}}
\resizebox{0.99\textwidth}{!}{\includegraphics*{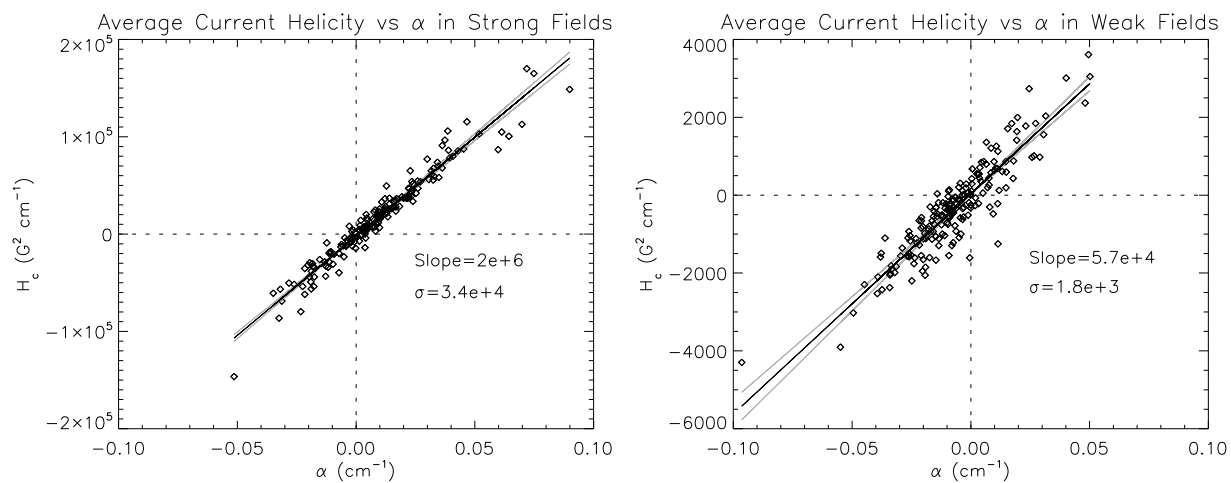}}
\end{center}
\caption{Scatter plots of the average photospheric current helicity $H_c$ against the average photospheric magnetic twist parameter $\alpha$ for weak (left) and strong (right) fields. Each point represents an active region. In each plot, the middle line represents the best fit using a linear regression model and the two grey lines represent  $2\sigma$ error bars. In each plot the values for the slope and $\sigma$ of the best-fit line from the linear regression analysis are quoted. The scaling of the vertical axes is not identical in the two plots.}
\label{fig:hc_alpha}
\end{figure}

\begin{figure} 
\begin{center}
%\resizebox{0.5\textwidth}{!}{\includegraphics*{Khel_hmi_gong.ps}}
\resizebox{0.5\textwidth}{!}{\includegraphics*{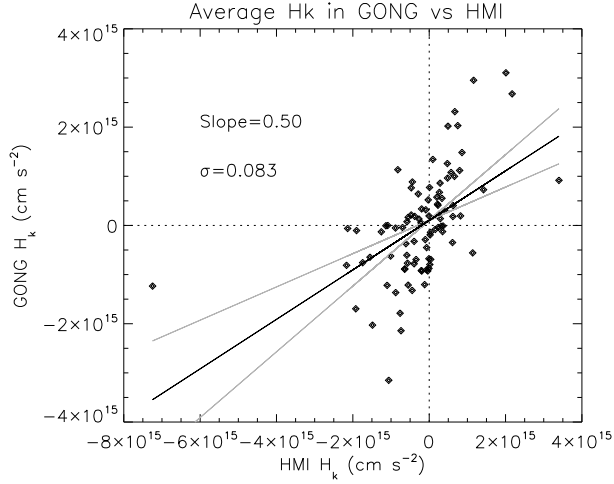}}
\end{center}
\caption{Scatter plots of the overlapping average subsurface kinetic helicity $H_k$ from GONG (2006-2012) against the  average subsurface kinetic helicity $H_k$ from HMI (2006-2012). Each point represents an active region. The middle line represents the best fit using a linear regression model and the two grey lines represent  $2\sigma$ error bars. The values for the slope and $\sigma$ of the best-fit line from the linear regression analysis are quoted.}
\label{fig:hk_gong_hmi}
\end{figure}

\begin{figure} 
\begin{center}
%\resizebox{0.99\textwidth}{!}{\includegraphics*{kh_alpha_hc_timerestricted_gong.ps}}
\resizebox{0.99\textwidth}{!}{\includegraphics*{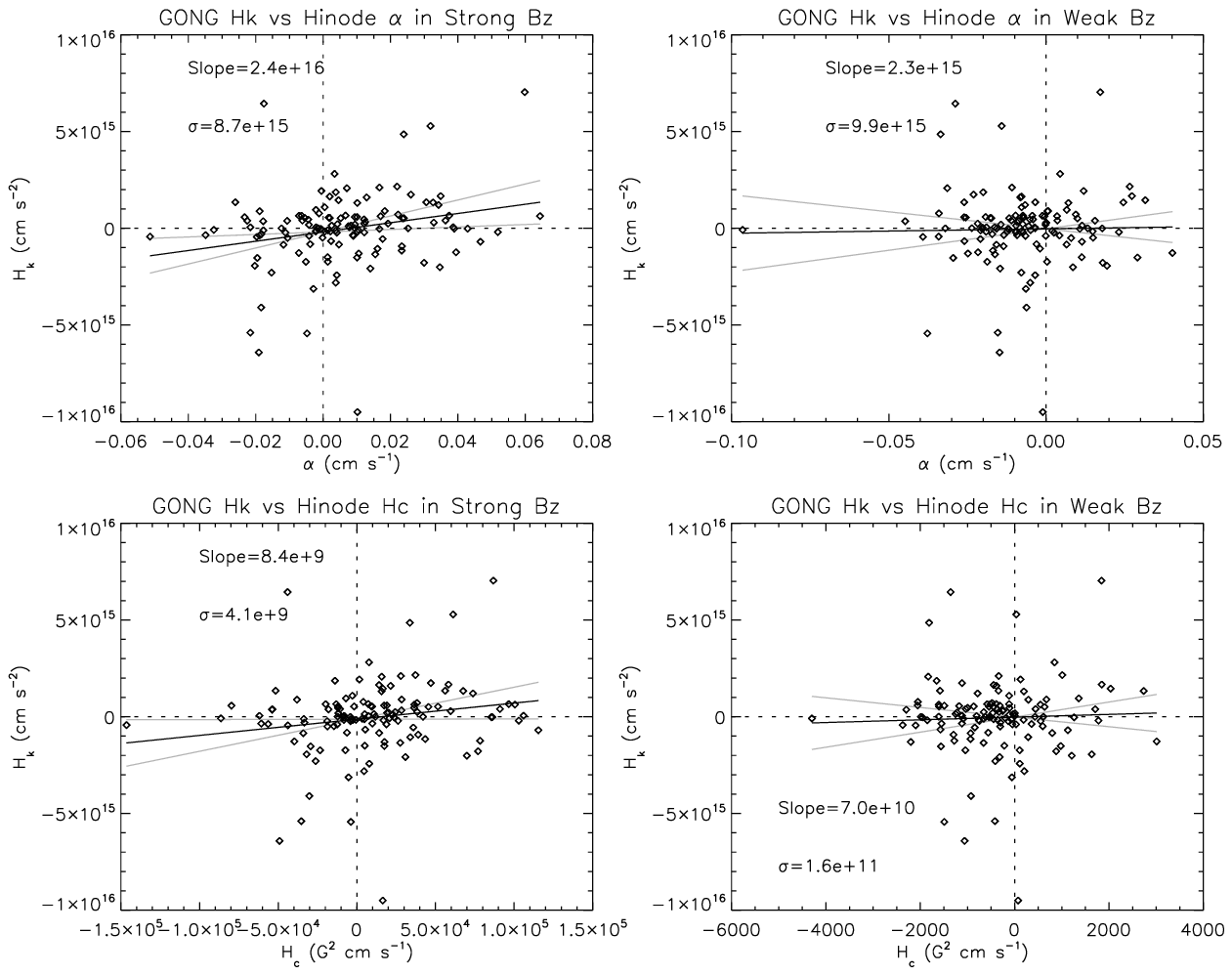}}
\end{center}
\caption{Scatter plots of the average subsurface kinetic helicity $H_k$ from GONG (2006-2012) data against the average photospheric magnetic twist parameter $\alpha$ (top) and the average photospheric current helicity $H_c$ (bottom)  for strong (left) and weak (right) fields. Each point represents an active region. In each plot, the middle line represents the best fit using a linear regression model and the two grey lines represent  $2\sigma$ error bars. In each plot the values for the slope and $\sigma$ of the best-fit line from the linear regression analysis are quoted. The scaling of the vertical axes is not identical in all plots.}
\label{fig:kh_gong}
\end{figure}

\begin{figure} 
\begin{center}
%\resizebox{0.99\textwidth}{!}{\includegraphics*{kh_alpha_hc_timerestricted_hmi.ps}}
\resizebox{0.99\textwidth}{!}{\includegraphics*{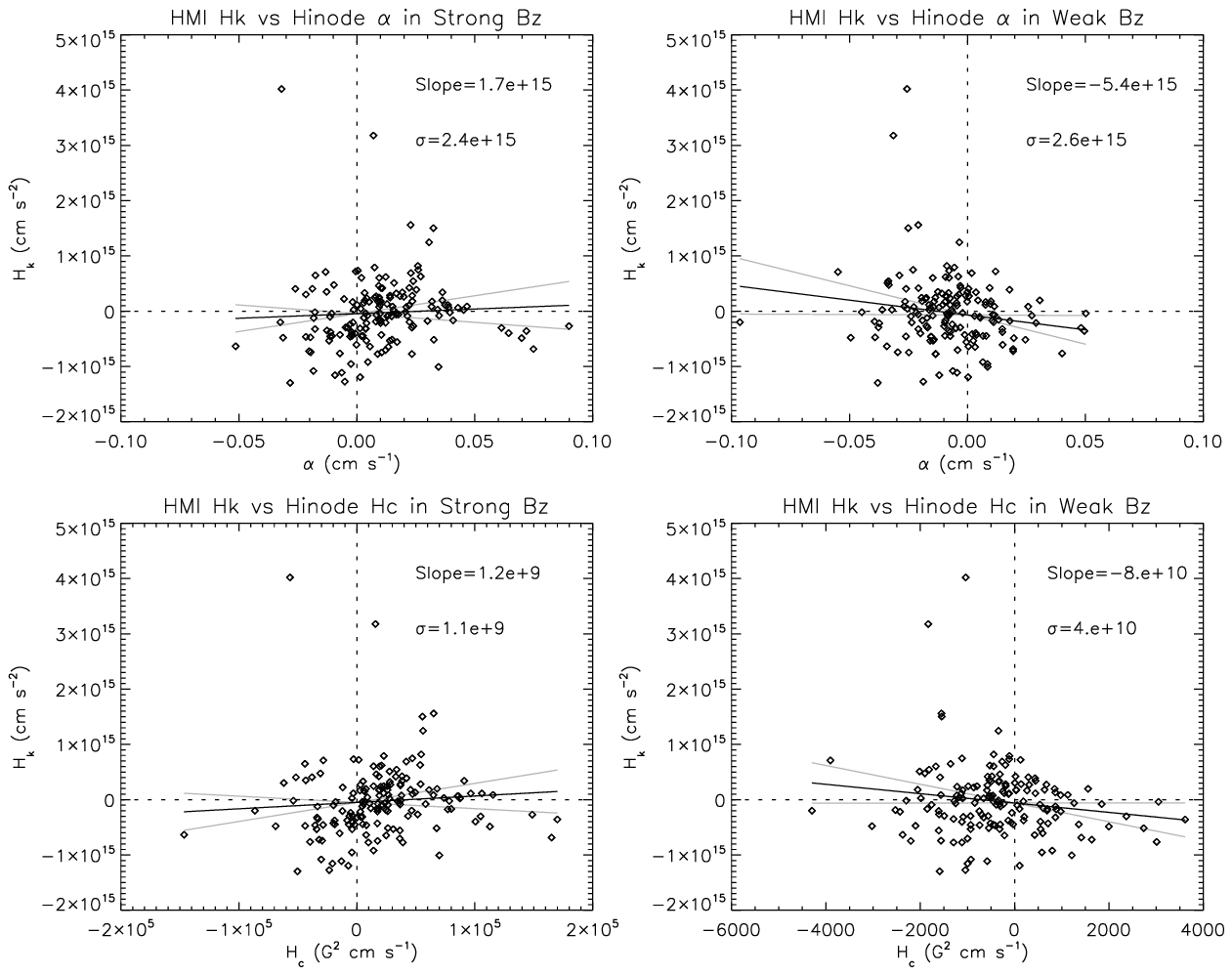}}
\end{center}
\caption{Scatter plots of the average subsurface kinetic helicity $H_k$ from HMI (2010-2013) data against the average photospheric magnetic twist parameter $\alpha$ (top) and the average photospheric current helicity $H_c$ (bottom)  for strong (left) and weak (right) fields. Each point represents an active region. In each figure, the middle line represents the best fit using a linear regression model and the two dashed lines represent  $2\sigma$ error bars.In each plot the values for the slope and $\sigma$ of the best-fit line from the linear regression analysis are quoted. The scaling of the vertical axes is not identical in all plots.}
\label{fig:kh_hmi}
\end{figure}

Figure~\ref{fig:hel_lat} shows plots of average twist parameter $<\alpha >$, average current helicity $<H_c>$, and average subsurface kinetic helicity $<H_k>$, all against heliographic latitude. Since we will discuss only averaged helicity values and not the helicity values associated with individual pixels, we will henceforth drop the $<>$ symbols and simply denote average twist parameter, current helicity and kinetic helicity values by $\alpha$, $H_c$ and $H_k$. For each region the average $\alpha$ and $H_c$ values are calculated by including all pixels in a magnetogram and by averaging the results over all magnetograms of the region.

We study weak and strong fields separately for the following reason. Zhang~(2006), Hao and Zhang~(2011) and Gosain et al.~(2013) found distinctly different patterns in weak and strong fields. In two examples, Hao and Zhang~(2011) found magnetic twist and current helicity of opposite sign in the umbra and penumbra, with the umbral field determining the overall sign of the active region. For each of their active regions, Zhang~(2006) and Gosain et al.~(2013) divided the field values into weak (100~G~$<|B_z|<$~500~G) and strong ($|B_z|>$~1000~G) classes and found that the two classes behaved differently. We adopt this approach so that our results can easily be compared to theirs. Thus each active region is represented by a single data point (194 points in total) in each plot in Figure~\ref{fig:hel_lat}.

The Hinode/SOT is a scanning instrument, and the pixel size in the direction of the raster is not identical to the pixel size along the slit. Also the projection angle varies as function of position on the disk, and this changes the spatial scale of the derivatives in Equations~(\ref{eq:alpha},\ref{eq:Hc}). These effects are accounted for in calculating $\alpha$ and $H_c$. Because of the high sensitivity and spatial resolution of the Hinode/SOT measurements, useful images could be taken of regions quite close to the limb, allowing much data to be included in the calculations. On the other hand, the Hinode/SOT is not a synoptic full-disk magnetograph and it has a limited field of view. The telescope points to different parts of the solar disk at different times so that the number of images collected per region varies.

The GONG and HMI Dopplergraphs nominally observe continuously with regular cadence, whereas Hinode observes different parts of the solar disk at different times. Pevtsov et al.~(1994) found that the characteristic decay time of a helicity pattern in an active region was about 27 hours. Therefore, for consistency, we derive average $H_k$ values based only on GONG or HMI measurements taken within a day of the nearest Hinode/SOT vector magnetogram of the region. In Figure~\ref{fig:hel_lat} the GONG (2006-2012) and HMI (2010-2013) $H_k$ values are plotted separately, and the Hinode average $\alpha$ and $H_c$ results for strong and weak fields are also plotted separately. Since we have 194 regions, and we calculate $H_c$ and $\alpha$ separately for the weak and strong fields in each region, every plot in Figure~\ref{fig:hel_lat} contains 194 data points. It is not practical to derive separate $H_k$ results for strong and weak fields because of the low spatial resolution of the $H_k$ data product and also because of the vertical spatial separation between the photosphere and the subsurface depths from which the $H_k$ data derive (see Section~2).

Clear patterns are visible in the top two plots of Figure~\ref{fig:hel_lat}. The GONG and HMI data sets both indicate hemispheric biases in $H_k$, tending to be positive in the southern hemisphere and negative in the northern hemisphere. The middle- and bottom-left plots of Figure~\ref{fig:hel_lat} likewise show that $\alpha$ and $H_c$ have the same bias for strong fields. The linear regression best fits are overplotted on the scatter plots with $2\sigma$ error bars. The linear regression `slope' coefficient indicated in each plot is the slope of the best-fit line. We shall henceforth refer to these linear regression coefficients as `slopes'. The $2\sigma$ error bars show the significance of the hemispheric biases. (The $2\sigma$ error bars represent 95\% confidence intervals, assuming that the results have normal distribution.) The matching hemispheric biases shown in the top two plots and the middle- and bottom-left plots are statistically significant because of the large size of the slope relative to $\sigma$. The middle- and bottom-right plots of Figure~\ref{fig:hel_lat} indicate that $\alpha$ and $H_c$ for weak fields have small hemispheric biases of opposite sign (negative in the southern hemisphere and positive in the northern hemisphere) but that these biases are not significant because of the large size of $\sigma$ relative to the slope.

Figure~\ref{fig:masterplot} plots the time variation of the patterns shown in Figure~\ref{fig:hel_lat}. Annual data bins were derived and linear regression calculations were performed for each helicity parameter against latitude for each year. The slope of each annual linear regression fit was recorded along with the slopes of the corresponding $2\sigma$ error bars. The results are plotted in Figure~\ref{fig:masterplot}. The figure shows that the strong-field $\alpha$ and $H_c$ averages had negative hemispheric bias over most but not all years, as did the $H_k$ averages. Here negative hemispheric bias indicates positive helicity in the southern hemisphere and negative helicity in the northern hemisphere, referring to the fact that the best-fit line has a negative slope, a convention based on how the data are plotted. The year 2007, which marked the end of cycle 23, stands out as a time when all helicities, photospheric and subsurface, weak- and strong-field, had positive hemispheric biases, the opposite of the prevailing overall pattern. This result is consistent with Hao and Zhang's~(2011) finding that late-cycle 23 active regions tended to have positive hemispheric helicity bias whereas early-cycle 24 regions tended to have negative hemispheric bias. In all other years the strong-field $\alpha$ and $H_c$ tended to have negative hemispheric bias, with the possible exception of 2009 when the activity level was unusually low (e.g., Wang et al.~2009). The subsurface $H_k$ does not appear to have had a hemispheric bias during that year either. During the remainder of the time period the annual average strong-field magnetic and current helicities and subsurface kinetic helicity had a negative hemispheric bias.

The annual average $\alpha$ and $H_c$ for weak fields initially followed the basic hemispheric trends of the other parameters (positive hemispheric bias in 2007, negative in 2008) but then did not show pronounced biases over the remainder of the time period, until perhaps 2013 when a sizable positive hemispheric bias emerged. The results therefore support Zhang et al.'s~(2010) conclusion that the hemispheric helicity bias can reverse at the beginnings and ends of activity cycles (here in 2007, at the end of cycle 23) but also agrees with Gosain et al.~(2013) in showing no overall reversal of bias at the beginning of cycle 24, and hints of positive hemispheric biases in the weak-field $\alpha$ and $H_c$.

The overall appearance of Figure~\ref{fig:masterplot} is striking in that the shapes of the graphs are so similar, even though the relationships between the helicities is evidently complicated. To explore these relationships further we compare the helicities directly in Figures~\ref{fig:hc_alpha}-\ref{fig:kh_hmi}.

Comparing Equations~(\ref{eq:alpha},\ref{eq:Hc}), the quantities $H_c$ and $\alpha$ are clearly related: $H_c = \alpha B_z^2$. It is therefore no surprise that the measurements of these quantities are well correlated with each other. Figure~\ref{fig:hc_alpha} shows scatter plots of $H_c$ against $\alpha$ for weak (100~G~$<|B_z|<$~500~G) and strong ($|B_z|>$~1000~G) magnetic field strengths separately. Each active region is represented by a data point in both plots. The Pearson linear correlation coefficient $cc$ is a measure of the linear dependence between two variables with $-1\le cc\le 1$, with $cc=\pm 1$ indicating perfect positive or negative correlation and $cc=0$ no correlation. The correlations of the variables plotted in Figure~\ref{fig:hc_alpha} are very good but not perfect. The Pearson linear correlation coefficients are 0.91 and 0.97 for the weak and strong fields, respectively. The quantities are clearly closely related but there is a difference in weighting by vertical field strength between them (compare Equations~(\ref{eq:alpha},\ref{eq:Hc})). Also, different weighted averages of signed quantities can sometimes give results of opposite sign, and in a small minority of cases the average $\alpha$ and $H_c$ do indeed have opposite sign. Overall, though, $\alpha$ and $H_c$ are highly correlated for both weak and strong fields.

Figure~\ref{fig:hk_gong_hmi} shows a scatter plot of the values of $H_k$ estimated from GONG data against the values of $H_k$ estimated from HMI data, for the subset of active regions observed by both telescopes. The correlation between these estimates is not nearly as good as the correlations between $H_c$ and $\alpha$ shown in Figure~\ref{fig:hc_alpha}, having Pearson linear correlation coefficient 0.54. This implies that the correlations between $H_k$ and $H_c$ or $\alpha$ will be at least partly compromised by the lower quality of the determination of $H_k$. The scatter in Figure~\ref{fig:hk_gong_hmi} is due to a combination of factors, including the different times the GONG and HMI telescopes observed each region, the different spectral bands used, the different spatial resolutions. Furthermore, spherical harmonics are fitted to higher harmonic order to the HMI data than to the GONG data, which may lead to larger contributions from near the surface in the HMI data. Another source of difference is that the inversion kernels in the GONG and HMI ring pipelines have different spatial distributions.

Figures~\ref{fig:kh_gong} and \ref{fig:kh_hmi} summarize the relationship between the photospheric helicities, $\alpha$ and $H_c$, and the subsurface kinetic helicity $H_k$, for the GONG and HMI $H_k$ measurements, respectively. It is immediately clear from the figures that the correlation between these quantities is much weaker and less significant than those shown in Figure~\ref{fig:hc_alpha}. The Pearson correlation coefficients are low and insignificant: around 0.2 for the GONG $H_k$ against the strong-field $\alpha$ and $H_c$ and significantly smaller for the weak-field $\alpha$ and $H_c$. For the HMI $H_k$ measurements the correlation coefficients are all smaller than 0.2. The photospheric and subsurface helicities are therefore not significantly correlated on a region-by-region basis. Given that all three parameters exhibit similar hemispheric biases for strong fields in Figure~\ref{fig:hel_lat}, some sign of weak positive correlation between them in Figure~\ref{fig:kh_gong} is perhaps not surprising, but the lack of significant correlation agrees with the results of Gao et al.~(2009).

The linear regressions plotted in Figure~\ref{fig:kh_gong} show positive linear fits for the GONG $H_k$ against the strong-field $\alpha$ and $H_c$ while the weak-field comparisons show no significant bias. The $H_k$ results from HMI cover a different time period, which may explain some of the differences between Figures~\ref{fig:kh_gong} and \ref{fig:kh_hmi}. In Figure~\ref{fig:kh_hmi} the distribution of HMI $H_k$ against the strong-field $\alpha$ and $H_c$ shows no significant slope in the linear fit whereas the weak-field linear fit has a negative slope that is barely significant to $2\sigma$.  These results reflect the fact that in Figure~\ref{fig:masterplot} $H_k$ had a generally negative hemispheric bias during 2010-2013 while the weak-field $\alpha$ and $H_c$ had a slight positive hemispheric bias during this time. It may seem surprising that the strong-field $\alpha$ and $H_c$ had a significant negative hemispheric bias during 2010-2013 and $H_k$ also had a marked negative hemispheric bias during this time, yet there is no significant slope in the linear regression between these quantities in Figure~\ref{fig:kh_hmi}. This detail illustrates a general property of these data: that quite strong hemispheric biases in photospheric and subsurface helicities generally do not extend to a direct region-by region correlation between the subsurface flow helicities and helicities of the photospheric magnetic field. On the other hand, the subsurface and photospheric helicities do seem to be related to each other overall, as demonstrated by the graphs in Figure~\ref{fig:masterplot}. The work has therefore revealed a link between subsurface kinetic and photospheric magnetic and current helicities, but one that so far has only become visible in comparisons of highly averaged quantities. This comparison between subsurface helicities may, of course, have been significantly affected by the quite poor determination of $H_k$ from the helioseismic data as indicated by Figure~\ref{fig:hk_gong_hmi}.

\section{Temporal Helicity Variations for Selected Active Regions}
\label{sect:temporal}

\begin{figure} 
\begin{center}
%\resizebox{0.8\textwidth}{!}{\includegraphics*[53,401][408,645]{master_hmi_alpha_strong.ps}}
%\resizebox{0.8\textwidth}{!}{\includegraphics*[53, 442] [410, 645]{master_hmi_alpha_weak.ps}}
\resizebox{0.8\textwidth}{!}{\includegraphics*[53,401][408,645]{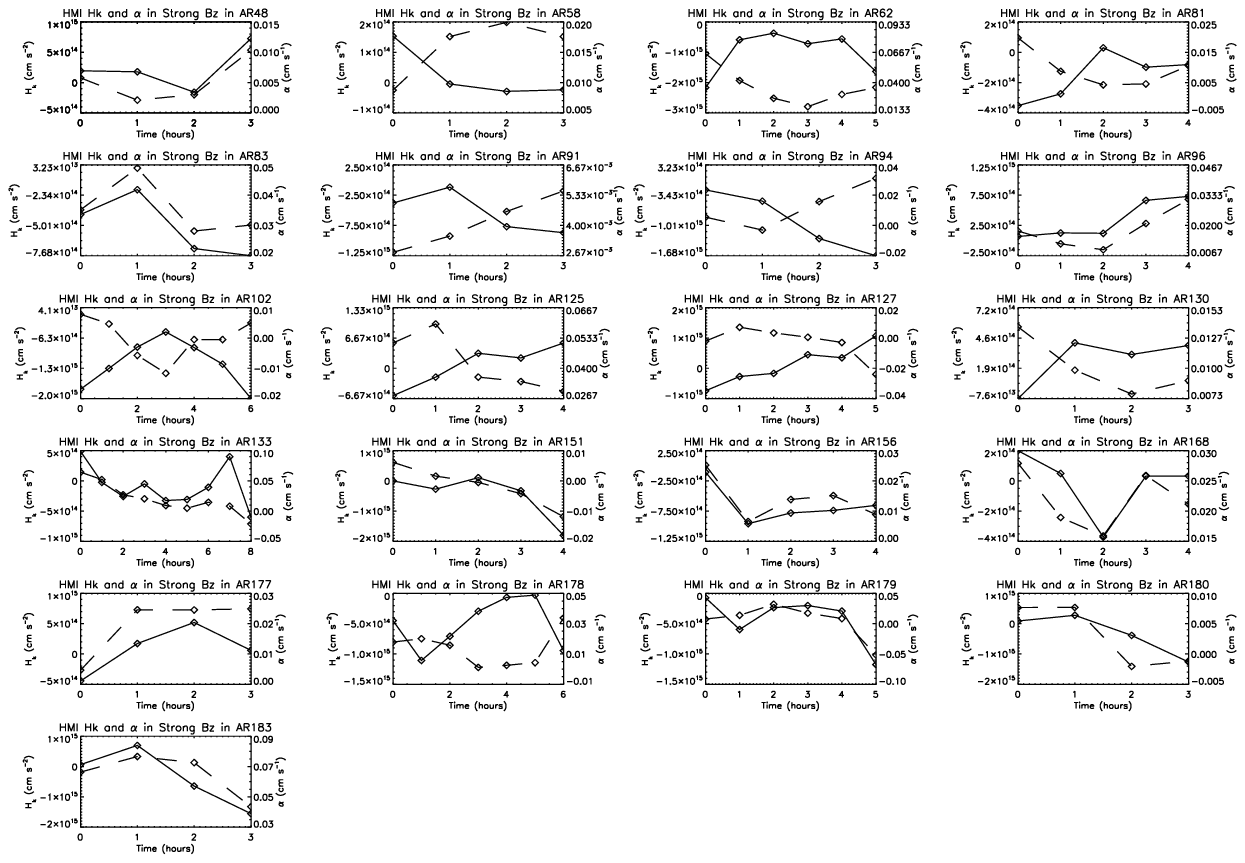}}
\resizebox{0.8\textwidth}{!}{\includegraphics*[53, 442] [410, 645]{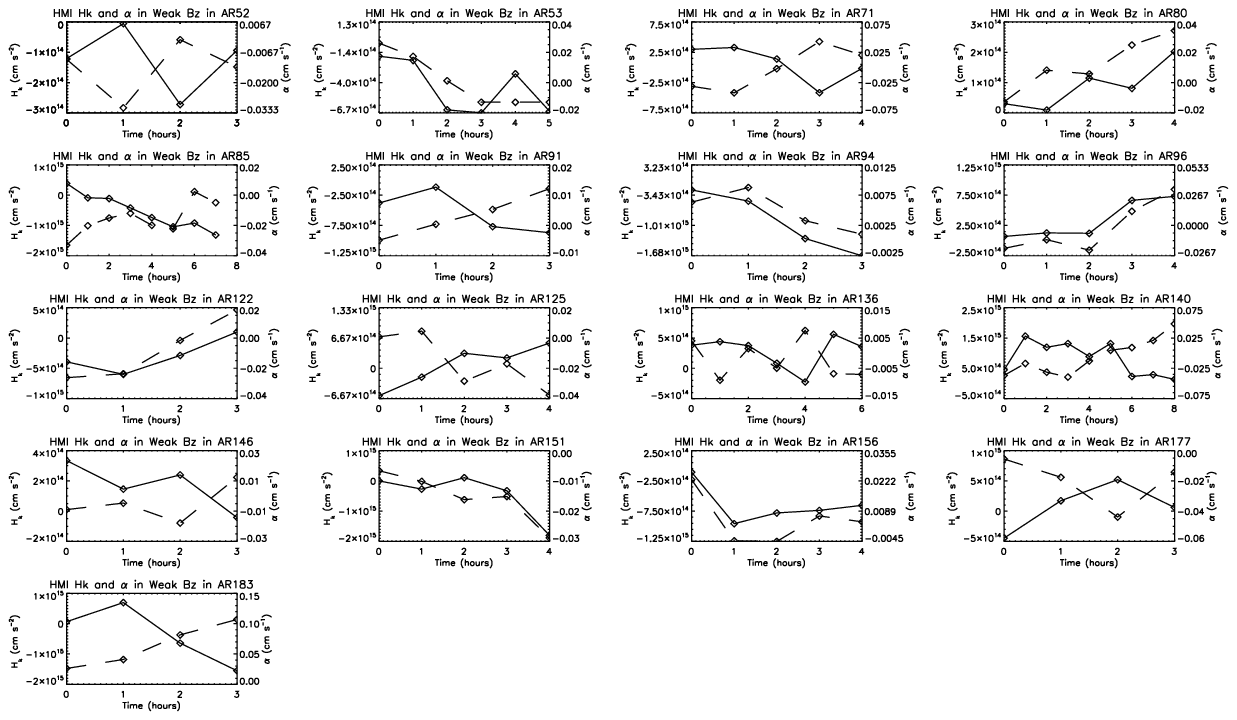}}
\end{center}
\caption{Plots of the average subsurface kinetic helicity $H_k$ (solid lines) from HMI (2010-2013) data and the average photospheric magnetic twist parameter $\alpha$ (dashed lines) in the strong and weak fields against time. Each $H_k$ data point represents the averaged $H_k$ from the HMI helioseismic data, and the Hinode data are binned according to the HMI ring pipeline time grid. The figure only includes plots representing regions for which good correlations or anti-correlations were found. Each plot is labeled by the number of the active region in our sample of 194 regions. Each region has Pearson linear correlation coefficient ($cc$) greater than .5 and a small associated p-value.}
\label{fig:kh_hmi_alpha_strong_time}
\end{figure}

\begin{figure} 
\begin{center}
%\resizebox{0.99\textwidth}{!}{\includegraphics*[53 ,401 ][409, 645]{master_hmi_hel_strong.ps}}
%\resizebox{0.99\textwidth}{!}{\includegraphics*[53, 483 ][408, 645]{master_hmi_hel_weak.ps}}
\resizebox{0.99\textwidth}{!}{\includegraphics*[53 ,401 ][409, 645]{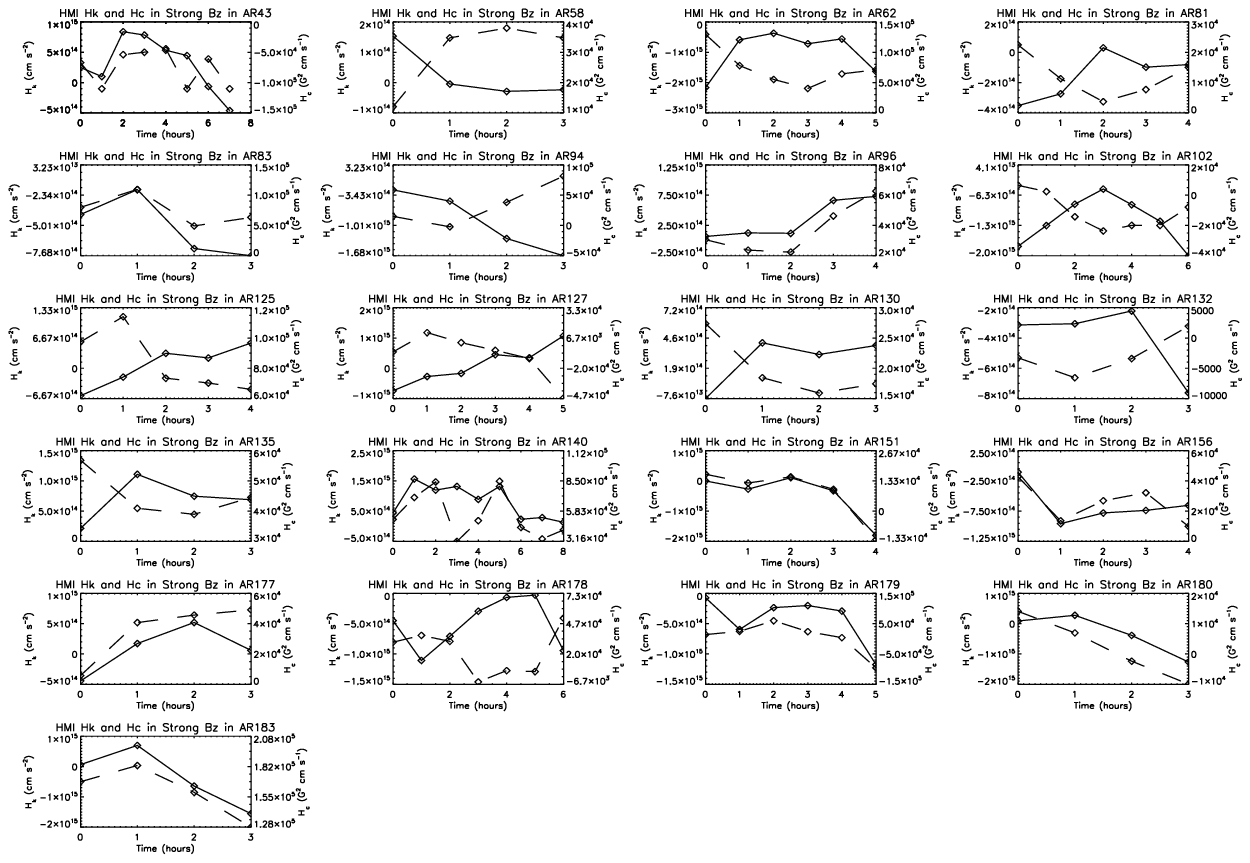}}
\resizebox{0.99\textwidth}{!}{\includegraphics*[53, 483 ][408, 645]{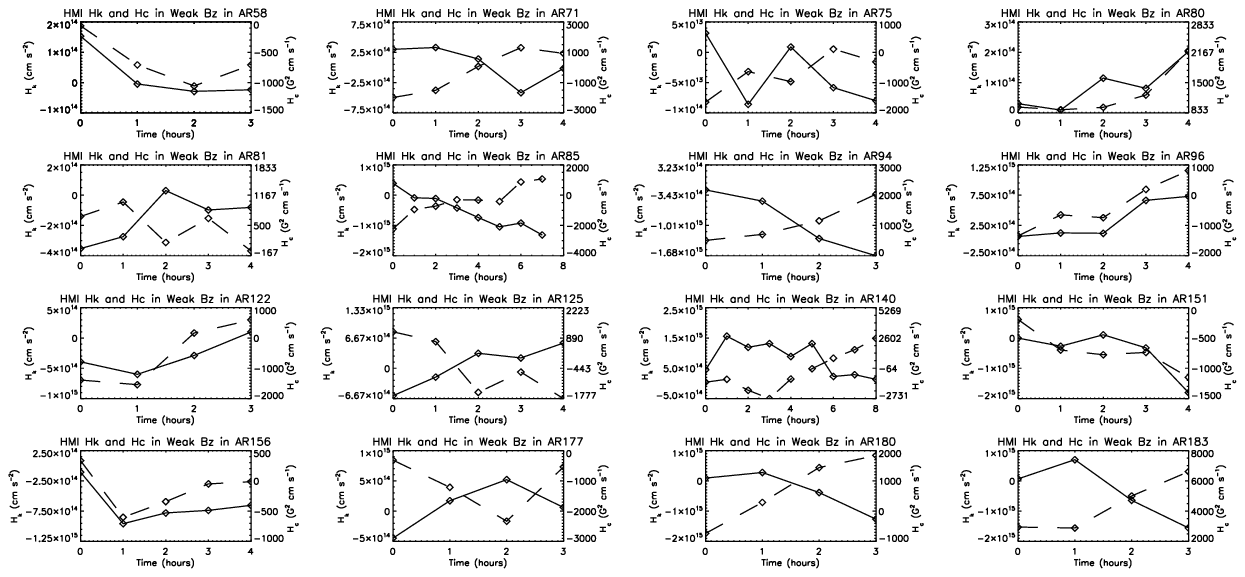}}
\end{center}
\caption{Plots of the average subsurface kinetic helicity $H_k$ (solid lines) from HMI (2010-2013) data and the average photospheric current helicity  $H_c$ (dashed lines) in the strong/weak fields against time. Each $H_k$ data point represents the averaged $H_k$ from the HMI helioseismic data, and the Hinode data are binned according to the HMI ring pipeline time grid. The figure only includes plots representing regions for which good correlations or anti-correlations were found. Each plot is labeled by the number of the active region in our sample of 194 regions. Each region has Pearson linear correlation coefficient ($cc$) greater than .5 and a small associated p-value.}
\label{fig:kh_hmi_hel_strong_time}
\end{figure}

\begin{figure} 
\begin{center}
%\resizebox{0.99\textwidth}{!}{\includegraphics*[52 ,500 ][404 ,640]{mastergongstrongalpha.ps}}
%\resizebox{0.99\textwidth}{!}{\includegraphics*[53 ,509 ][404, 640]{mastergongweakalpha.ps}}
\resizebox{0.99\textwidth}{!}{\includegraphics*[52 ,500 ][404 ,640]{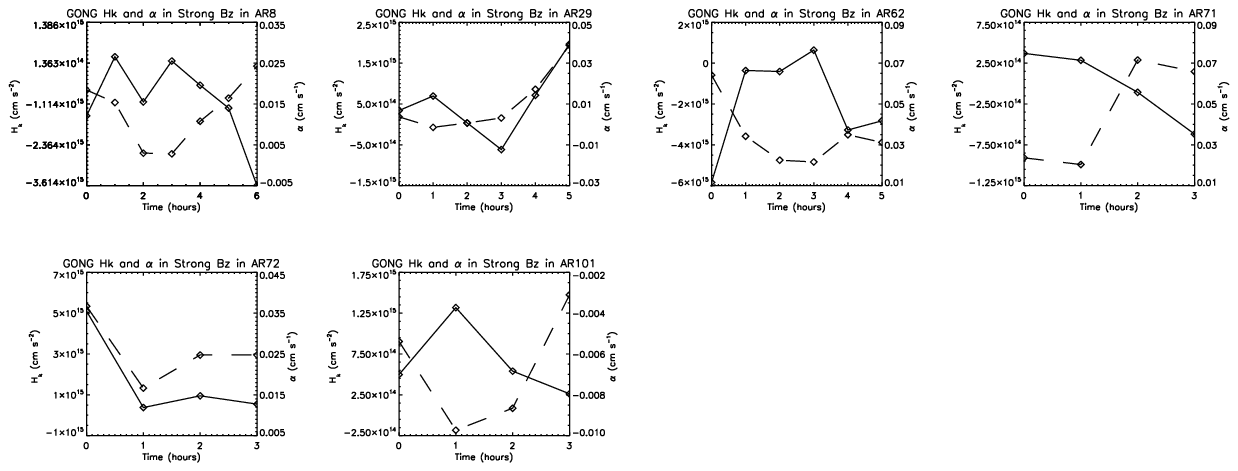}}
\resizebox{0.99\textwidth}{!}{\includegraphics*[53 ,509 ][404, 640]{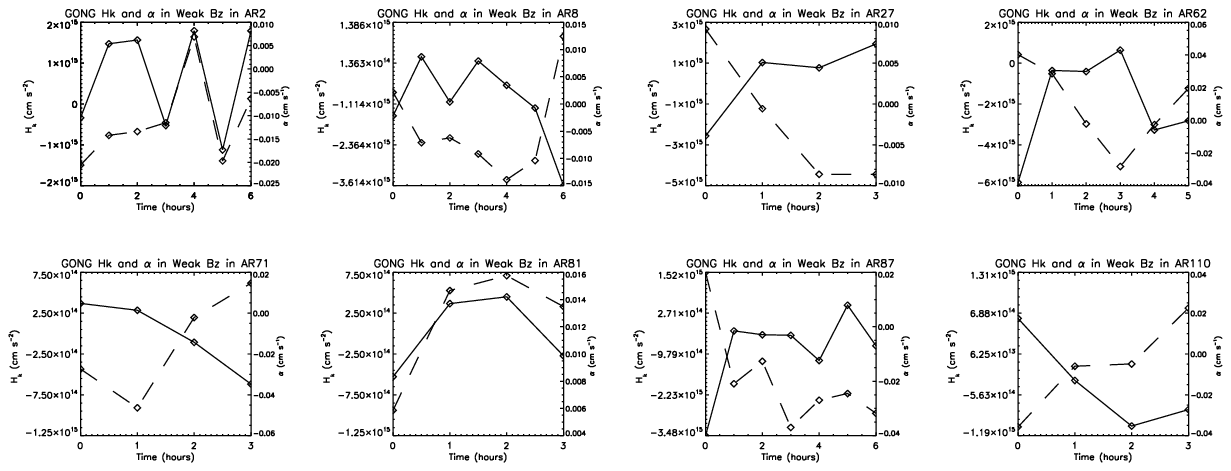}}
\end{center}
\caption{Plots of the average subsurface kinetic helicity $H_k$ (solid lines) from GONG (2006-2010) data and the average photospheric magnetic twist parameter $\alpha$ (dashed lines) in the strong and weak fields against time. Each $H_k$ data point represents the averaged $H_k$ from the GONG helioseismic data, and the Hinode data are binned according to the GONG ring pipeline time grid. The figure only includes plots representing regions for which good correlations or anti-correlations were found. Each plot is labeled by the number of the active region in our sample of 194 regions. Each region has Pearson linear correlation coefficient ($cc$) greater than .5 and a small associated p-value. }
\label{fig:kh_gong_alpha_strong_time}
\end{figure}

\begin{figure} 
\begin{center}
%\resizebox{0.99\textwidth}{!}{\includegraphics*[52 ,500 ][411, 640]{mastergongstronghc.ps}}
%\resizebox{0.99\textwidth}{!}{\includegraphics*[52 ,509 ][408, 640]{mastergongweakhc.ps}}
\resizebox{0.99\textwidth}{!}{\includegraphics*[52 ,500 ][411, 640]{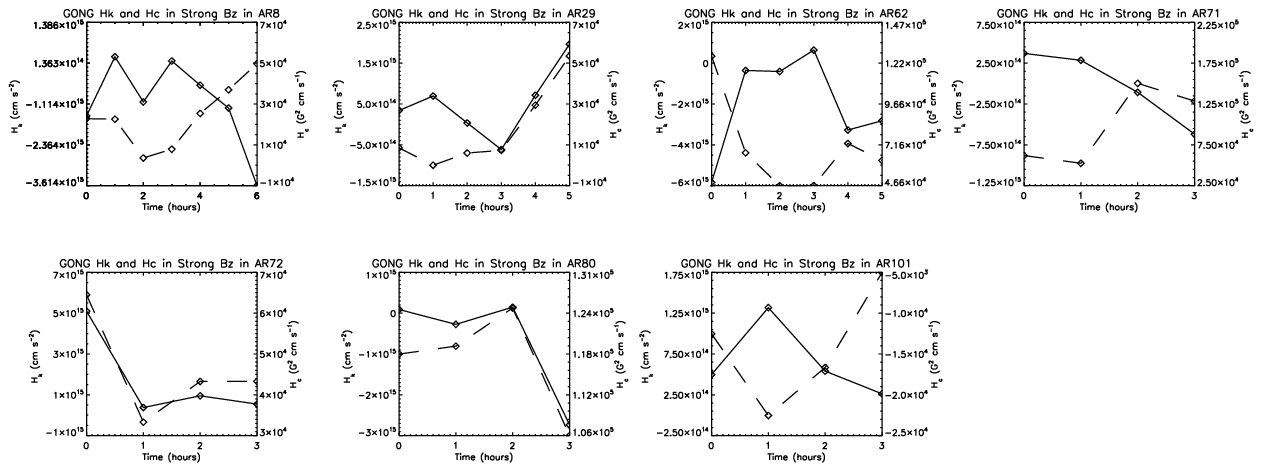}}
\resizebox{0.99\textwidth}{!}{\includegraphics*[52 ,509 ][408, 640]{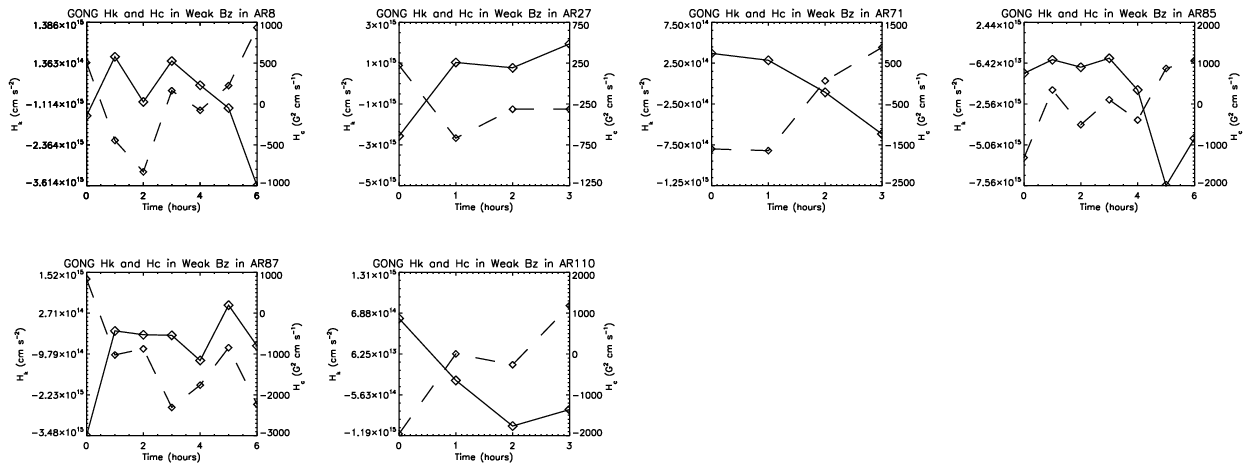}}
\end{center}
\caption{Plots of the average subsurface kinetic helicity $H_k$ (solid lines) from GONG (2006-2010) data and the average photospheric current helicity  $H_c$ (dashed lines) in the strong and weak fields against time. Each $H_k$ data point represents the averaged $H_k$ from the GONG helioseismic data, and the Hinode data are binned according to the GONG ring pipeline time grid. The figure only includes plots representing regions for which good correlations or anti-correlations were found. Each plot is labeled by the number of the active region in our sample of 194 regions. Each region has Pearson linear correlation coefficient ($cc$) greater than .5 and a small associated p-value.}
\label{fig:kh_gong_hel_strong_time}
\end{figure}

\begin{table}
\scriptsize
\caption{Statistics for regions with significant correlations between time profiles of $H_k$ and of $H_c$ or $\alpha$ - see Figures~\ref{fig:kh_hmi_alpha_strong_time}-\ref{fig:kh_gong_hel_strong_time}. The columns divide the statistics into cases with $H_k$ data from GONG and from HMI, and with $H_k$ the same sign as or opposite sign to $H_c$ or $\alpha$. With $\theta$ denoting latitude, the condition $\theta H_k <0$ identifies regions whose kinetic helicities follow the usual hemispheric rule (see text). The variable $cc$ is the Pearson linear correlation coefficient. Numbers of cases where $H_k$ has the same/opposite sign as $H_c$ or $\alpha$ on average are shown, indicating how many of these cases have positively ($cc>0$) or negatively ($cc<0$) correlated time profiles of $H_k$ compared to $H_c$ or $\alpha$. The table shows that cases with like-/opposite-sign subsurface and photospheric helicities tend to have positively/negatively correlated subsurface and photospheric helicity time profiles. This holds whether or not $H_k$ follows the hemispheric rule, $\theta H_k <0$.}\
\label{statstable}
\\
\begin{tabular}{l*{5}{c}r}
              & GONG $H_k$ same sign:  & GONG $H_k$ opposite sign:  & HMI $H_k$ same sign: & HMI $H_k$ opposite sign  \\
              &  \#$cc>0$/\# total &  \#$cc<0$/\# total  &  \#$cc>0$/\# total  &  \#$cc<0$/\# total  \\
\hline
$\alpha$ (Strong)  & 2/3 & 3/3  & 6/7 & 9/14   \\
$\alpha$ (Weak)         & 2/3 & 5/5 & 3/7 & 6/10 \\
$H_c$ (Strong)         & 3/3 & 4/4  & 5/8 & 8/13  \\
$H_c$ (Weak)      & 1/2 & 4/4  & 4/7 & 7/9  \\
$\alpha$ (Strong, $\theta H_k <0$) & 2/2 & 1/1  & 4/5 & 6/10   \\
$\alpha$ (Weak, $\theta H_k <0$)         & 1/2 & 2/2 & 1/4 & 5/9 \\
$H_c$ (Strong, $\theta H_k <0$)         & 3/3 & 1/1  & 3/5 & 5/9  \\
$H_c$ (Weak, $\theta H_k <0$)      & 1/2 & 1/1  & 2/4 & 6/8  \\
\hline
\end{tabular}
\end{table}

To explore further the relationship between the helicities associated with the subsurface flows and the photospheric magnetic fields and electric currents, we searched for cases in our sample where the temporal profiles of these quantities were reasonably well correlated. In particular, we rebinned the $H_c$ and $\alpha$ values on the daily $H_k$ temporal grid, compared the resulting temporal profiles of the three quantities, and selected cases where the Pearson linear correlation coefficient $cc\ge 0.5$ and the associated p-values were small ($\leq 0.1$). The p-value is the probability that such a good (or a better) linear correlation between the quantities could have been produced by chance alone. A small p-value therefore indicates that the correlation did not occur by chance, but represents a real link between the quantities. In total we found temporal correlations between $H_k$ and $\alpha$ for 14 regions observed by GONG and 38 observed by HMI, and  temporal correlations between $H_k$ and $H_c$ for 16 regions observed by GONG and 37 observed by HMI. The results are shown in Figures~\ref{fig:kh_hmi_alpha_strong_time}-\ref{fig:kh_gong_hel_strong_time} and are summarized in Table~\ref{statstable}. Figures~\ref{fig:kh_hmi_alpha_strong_time} and \ref{fig:kh_hmi_hel_strong_time} summarize the comparisons between the time-averaged magnetic and current helicities from Hinode and average helioseismic kinetic helicities from HMI. Figures~\ref{fig:kh_gong_alpha_strong_time} and \ref{fig:kh_gong_hel_strong_time} summarize the equivalent comparisons for the GONG helioseismic kinetic helicity data set. Note that there are differences between the $\alpha$ and $H_c$ plots. Though $\alpha$ and $H_c$ are related in theory (compare Equations~(\ref{eq:alpha},\ref{eq:Hc})) and in the data (see Figure~\ref{fig:hc_alpha}), these two parameters can behave quite differently in some cases. In the cases included in Table~\ref{statstable} we found a strong pattern relating the signs of the helicities and the signs of the linear correlation coefficients. In each subset of regions, we found both highly correlated and anti-correlated regions. Examples with subsurface and photospheric helicities of the same sign tended to have positively correlated temporal evolution in these helicities, and examples with subsurface and photospheric helicities of opposite sign tended to have anti-correlated temporal evolution in these helicities. In general, the two parameters are correlated when they have the same sign and are anti-correlated when they have opposite sign. This indicates that the unsigned magnetic and kinetic helicities are highly correlated in these regions. This result is consistent with the notion that whichever process was responsible for creating the helicities of like/opposite signs may also have governed the temporal changes in these helicities.

The bottom four rows of Table~\ref{statstable} show that the cases with negative/positive $H_k$ in the northern/southern hemisphere, i.e., those whose vertical kinetic helicities match the sign of the Coriolis force, do not behave significantly differently from the other cases. The numbers in the last four columns of Table~\ref{statstable} show the same patterns as the first four rows: subsurface and photospheric helicities of like/opposite sign tend to occur in cases with positively/negatively correlated temporal behavior in these quantities. The influence on the photospheric fields of the flows satisfying the hemispheric rule is indistinguishable from the influence of flows violating the hemispheric rule.

The majority of the regions included in Table~\ref{statstable} also featured subsurface and photospheric helicities of opposite sign: the 2nd and 4th columns of Table~\ref{statstable} generally contain larger numbers than the 1st and 3rd columns. In view of the results discussed earlier and summarized in Figures~\ref{fig:kh_gong} and \ref{fig:kh_hmi}, the strong-field comparisons are expected to be biased towards cases with same-sign helicities, at least for GONG data, and the strong-field comparisons are not expected to show much bias either way. The preference in Table~\ref{statstable} for subsurface and photospheric helicities having opposite sign therefore suggests that the cases with like-sign helicities were under-represented in Table~\ref{statstable}. As discussed in Section~\ref{sect:intro}, we might expect opposite-sign subsurface and photospheric helicities to be associated with the $\alpha$-effect on weak, widely-distributed fields and like-sign helicities to be associated with the $\Sigma$-effect on isolated flux tubes. If this is true, then it may be the coarse spatial resolution of the helioseismic kinetic helicity data, that makes it difficult to detect good correlations associated with the localized eddies responsible for the $\Sigma$-effect photospheric helicity changes than the more widely distributed $\alpha$-effect flow patterns. However, the overall pattern of positive/negative correlation between like-/opposite-sign helicities appears to be solar in origin.

\section{Conclusion}
\label{sect:conclusion}

Seeking evidence of a physical link between helicity patterns in the fluid-dominated solar interior and the magnetically-dominated solar atmosphere, we have collected observational data for the subsurface kinetic helicity $H_k$ and for the photospheric current helicity $H_c$ and twist parameter $\alpha$ for the period 2006-2013.

We compared the data for these three helicity parameters, and found that $H_c$ and $\alpha$ obey the usual hemispheric helicity rule for strong fields, and that $H_k$ has a similar hemispheric bias as shown in Figure~\ref{fig:hel_lat}. Although there was no significant evidence that the values for $H_c$ and $\alpha$, averaged for each region in space and time, correlate with the average $H_k$ values for the set of 194 regions overall (Figures~\ref{fig:kh_gong} and \ref{fig:kh_hmi}), the annual average graphs of $H_k$, $H_c$ and $\alpha$ (Figure~\ref{fig:masterplot}) showed striking similarities and indicated similar annual variations in the parameters. Further evidence of a physical link between the subsurface and photospheric helicities was found in a subset of 77 cases where the temporal profiles of the active regions' $H_c$ and $\alpha$ correlated well with the $H_k$ temporal profiles (Figures~\ref{fig:kh_hmi_alpha_strong_time}-\ref{fig:kh_gong_hel_strong_time}). The signs of the Pearson linear correlation coefficients between the temporal profiles generally matched the sign relationship between the profiles themselves, pointing to a common physical cause of the subsurface and photospheric helicities at the beginning of the time series and their subsequent variations.

Referring to Section~\ref{sect:intro}, the $\Sigma$-effect is characterized by right-/left-handed fluid motions ultimately producing right-/left-handed magnetic field structure, whereas the $\alpha$-effect is consistent with left-/right-handed field structure being produced by these motions. In our data set there is no such bias overall: subsurface fluid motions with a given sign of kinetic helicity appear to correspond to photospheric field structures of the same and of opposite handedness in approximately equal numbers. However, the evidence of temporal relationships between the subsurface kinetic and photospheric magnetic and current helicities suggests a physical link between the subsurface velocity fields and the photospheric magnetic fields and electric currents of a form similar to the $\alpha$-effect, the $\Sigma$-effect, or some combination of the two.

\acknowledgements{}

We thank the referee for constructive and helpful comments that helped to clarify the paper. D.S. carried out this work through the National Solar Observatory Research Experiences for Undergraduate (REU) site program, which is co-funded by the Department of Defense in partnership with the National Science Foundation REU Program. This work was supported by NSF/SHINE Award No. 1062054 to the National Solar Observatory. This work utilizes data obtained by the NSO Integrated Synoptic Program (NISP), managed by the National Solar Observatory, which is operated by AURA, Inc. under a cooperative agreement with the National Science Foundation. The data were acquired by instruments operated by the Big Bear Solar Observatory, High Altitude Observatory, Learmonth Solar Observatory, Udaipur Solar Observatory, Instituto de Astrof\'{\i}sica de Canarias, and Cerro Tololo Interamerican Observatory. Hinode is a Japanese mission developed and launched by ISAS/JAXA, collaborating with NAOJ as a domestic partner, NASA and STFC (UK) as international partners. Scientific operation of the Hinode mission is conducted by the Hinode science team organized at ISAS/JAXA. This team mainly consists of scientists from institutes in the partner countries. Support for the post-launch operation is provided by JAXA and NAOJ (Japan), STFC (U.K.), NASA, ESA, and NSC (Norway).

\end{document}